\documentclass{ws-ijmpa}
\usepackage[super,compress]{cite}
\usepackage{graphicx}
\usepackage{xcolor}
\begin{document}
\markboth{Campbell {\it et al.}}{Seeking the Casimir Energy}

%\catchline{}{}{}{}{}
%
%%%%%%%%%%%%%%%%%%%%%%%%%%%%%%%%%%%%%%%%%%%%%%%%%%%%%%%%%%%%%%%%%%%%
%---------------------- editing macros ------------------

\newcommand{\as}[1]{\textcolor{cyan}{[\textit{Abhishek}: #1]}}
%---------------------- editing macros ------------------
\title{Seeking the Casimir Energy
}

\author{David K. Campbell
}

\address{Department of Physics, Boston University, Boston, Massachusetts, 02215\\
USA\\
dkcampbe@bu.edu}

\author{Ian Bouche}

\address{Department of Physics, Boston University, Boston, Massachusetts, 02215\\
USA\\
ianbo@bu.edu}

\author{Abhishek Som}

\address{Department of Physics, Boston University, Boston, Massachusetts, 02215\\
USA\\
asom5@bu.edu}

\author{David J. Bishop}

\address{Materials Science and Engineering Division, Boston University, Boston, Massachusetts, 02215\\
USA\\
djb1@bu.edu}
\maketitle

\begin{history}
\received{\today}
\revised{\today}
\end{history}

\begin{abstract}
Since its first description in 1948, the {\it Casimir effect} has been studied extensively. Standard arguments for its existence hinge on the elimination of certain modes of the electromagnetic field because of the boundary conditions in the Casimir cavity. As such, it has been suggested that the ground state energy of the vacuum within the cavity may be reduced compared to the value outside. Could this have an effect on physical phenomena within the cavity? We study this {\it Casimir energy} and probe whether the critical temperature $T_c$ of a superconductor is altered when it is placed in the cavity. We do not detect any change in $T_c$ larger than 12 microKelvin, but theoretically expect a change on the order of 0.025 microKelvin, roughly 1000 times lower than our achieved sensitivity.

\end{abstract}

\ccode{PACS numbers:}

%\tableofcontents

\section{Introduction}	
The Casimir force was first derived in 1948 by calculating
the van der Waals force using retarded potentials\cite{cas1}.
This force is a purely quantum mechanical force that
arises between two plates even when they are not electrically
charged. Classically, there is no force on the plates.
However, due to quantum fluctuations and the freezing
out of long-wavelength electromagnetic modes, there
is a net pressure exerting an attractive force. Experimentally,
the effect has been seen using a number of
microscale systems and devices involving differing metallic conductivity (see for instance Refs.~\refcite{lam78,moh98}.)

Given that decreasing metallic conductivity increases the Casimir force, one would expect that a superconducting material should have much stronger Casimir forces. However, the Casimir effect averages the conductivity of
the material over very large energy scales, while the
superconducting gap is relevant only for the far infrared. 
Therefore, while the effect of superconductivity is very
large (100\%) on the DC conductivity, it is negligible and
unmeasurable if averaged over the typical energy scales
found in a Casimir cavity. For this reason, one cannot see an
effect on the measured Casimir force at the transition
temperature $T_c$.

\begin{figure}[h]
\centerline{\includegraphics[width=12.0cm]{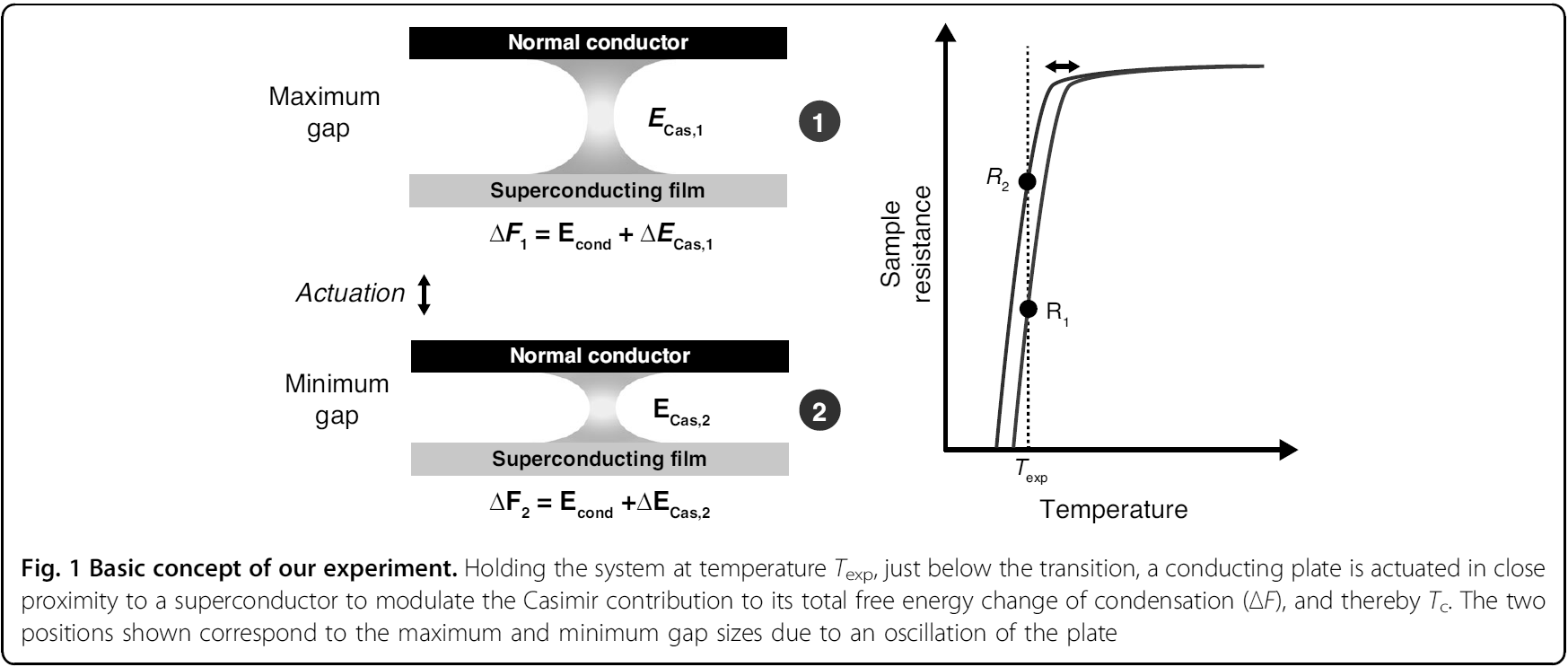}}
\caption{A schematic illustration of (left panel) our variable width Casimir cavity with one face being superconducting and the other normal metal; and (right panel) showing the effect on the superconducting $T_c$ as we vary the distance between the walls of the cavity.
\label{fig1}}
\end{figure}

In a recent article\cite{diego2020}\negthinspace, building on earlier work by Bimonte and collaborators\cite{bimonte2005}\negthinspace, we have described a microelectromechanical system (MEMS) that would in principle allow for the measurement of the change in critical temperature of a superconductor in a Casimir cavity: a superconductor is placed inside a metallic Casimir cavity by depositing a superconducting material on one of the metallic plates in the cavity and measuring the change in the $T_c$ of the material as shown in Fig. \ref{fig1}. We refer to Ref. \refcite{diego2020} for details.

\section{Experimental Results and Theoretical Interpretation}
There are a number of experimental challenges to measure
by the effect shown in Fig. \ref{fig1}. According to calculations performed in Ref. \refcite{bimonte2005}, the spacing of the Casimir cavity should be in the range of a few nanometers to 100 nm, and the film thickness should be on the order of 10 nm. These characteristics place serious constraints on the choice of materials, many of which tend to ball up and form islanded microstructures when thin. However,
evaporating onto cryogenically cooled surfaces allows for the quenched condensation of the material, forming very smooth, amorphous films\cite{diego2020}\negthinspace. For this reason, we used an {\it in situ} deposition method in which the superconducting film is deposited at the chip-scale, below the superconducting transition temperature. This fab-on-a-chip methodology is explained further in the methods section of Ref. \refcite{diego2020}. It is with this quenched-condensed thin film (which serves as one half of a tunable Casimir cavity) that we are able to probe changes in the Casimir energy. The results of sweeping through the transition in temperature are shown in FIg. 2 below.

\begin{figure}[h]
\centerline{\includegraphics[width=12.0cm]{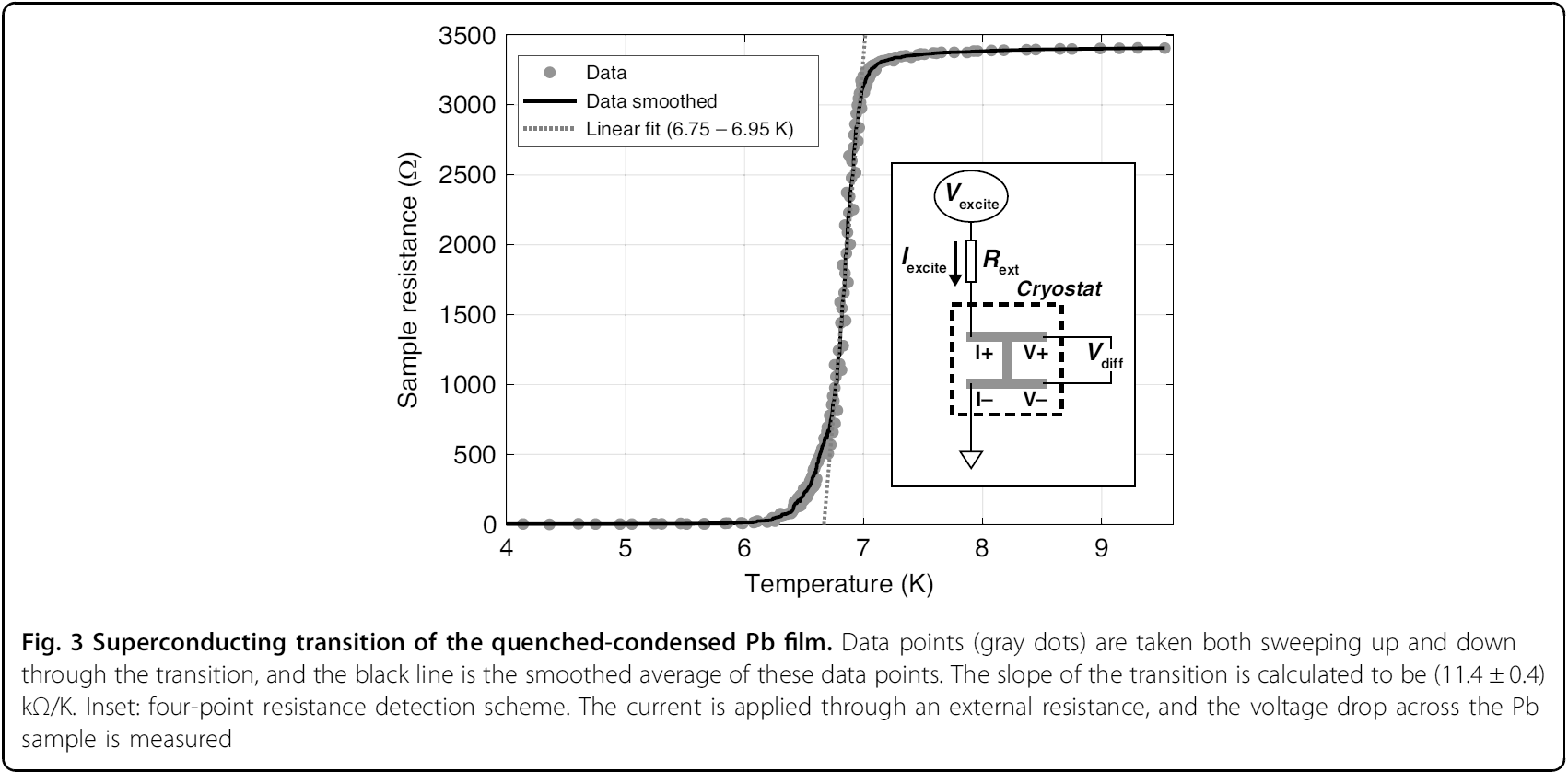}}
\caption{Data taken from sweeping up and down in temperature through the transition showing that a linear fit is an accurate representation of the data. From Fig 3 of Ref. \protect\refcite{diego2020}) 
\label{fig2}}
\end{figure}

\begin{figure}[h]
\centerline{\includegraphics[width=12.0cm]{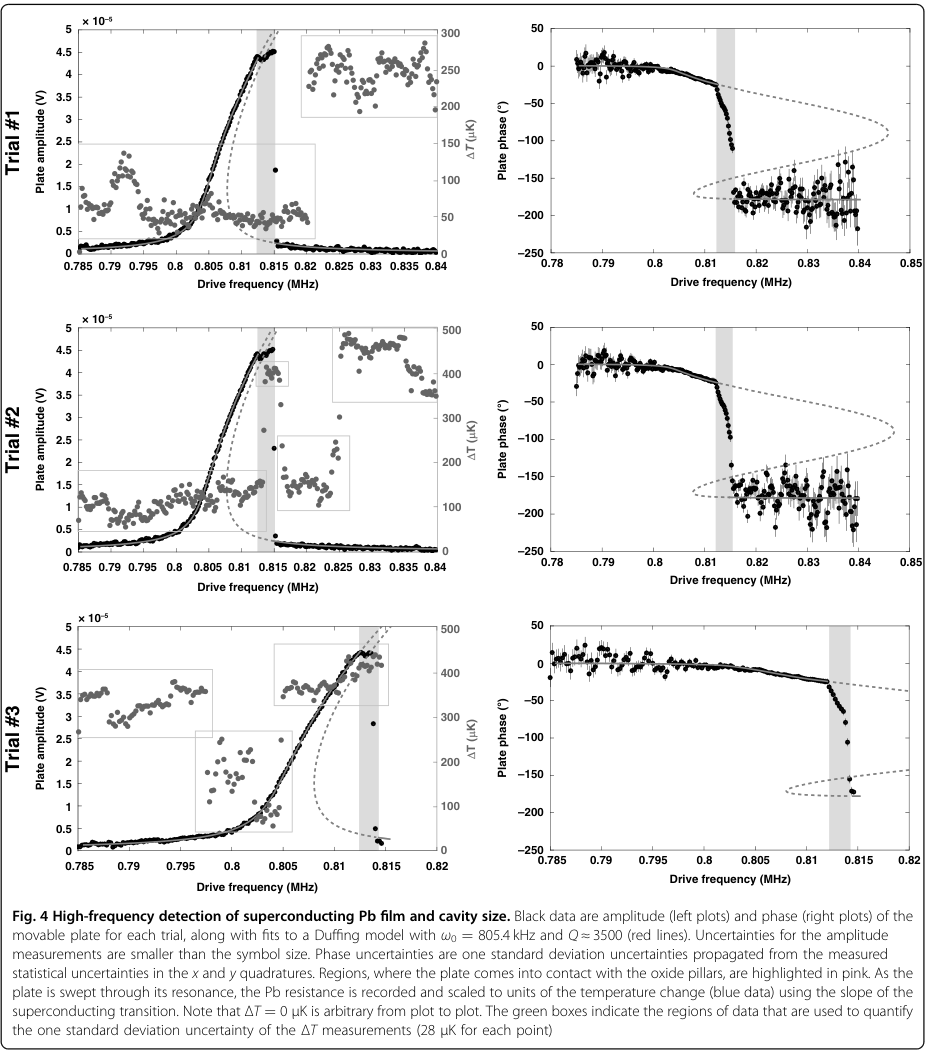}}
\caption{Fig 4 from  Ref. \protect\refcite{diego2020}. The high frequency detection of the superconducting lead film and the size of the Casimir cavity in two trial runs of our experiment. The black data in the left plots are the amplitude of the movable plate for each trial while the black data in the right plots are the phase. The amplitude and phase are plotted as functions of  the drive frequency in MHz. We are able to separate amplitude and phase because of our use for frequency dependent detection. Further details, including the color which could not be reproduced in this journal, are available in the original reference.
\label{fig3}}
\end{figure}

In figure \ref{fig3} below (taken from Ref \refcite{diego2020}), we display experimental data of the amplitude (left plots) and phase (right plots) of the movable plate. Regions where the plate comes into contact with parts of the MEMS apparatus are shown as shaded. As the plate is swept through the resonance, the resistance of the Pb is recorded and scaled to units of the temperature change (which is shown in the data points) using the slope of the superconducting transition. 
%In figure \ref{fig4{ below (also taken from Ref \refcite{diego2020})
\newpage

%In Ref ~\refcite{diego2020} we studied the high frequency detection of the superconducting PB file and the cavity size. 

\begin{figure}[h]
    \centerline{\includegraphics[width=10cm]{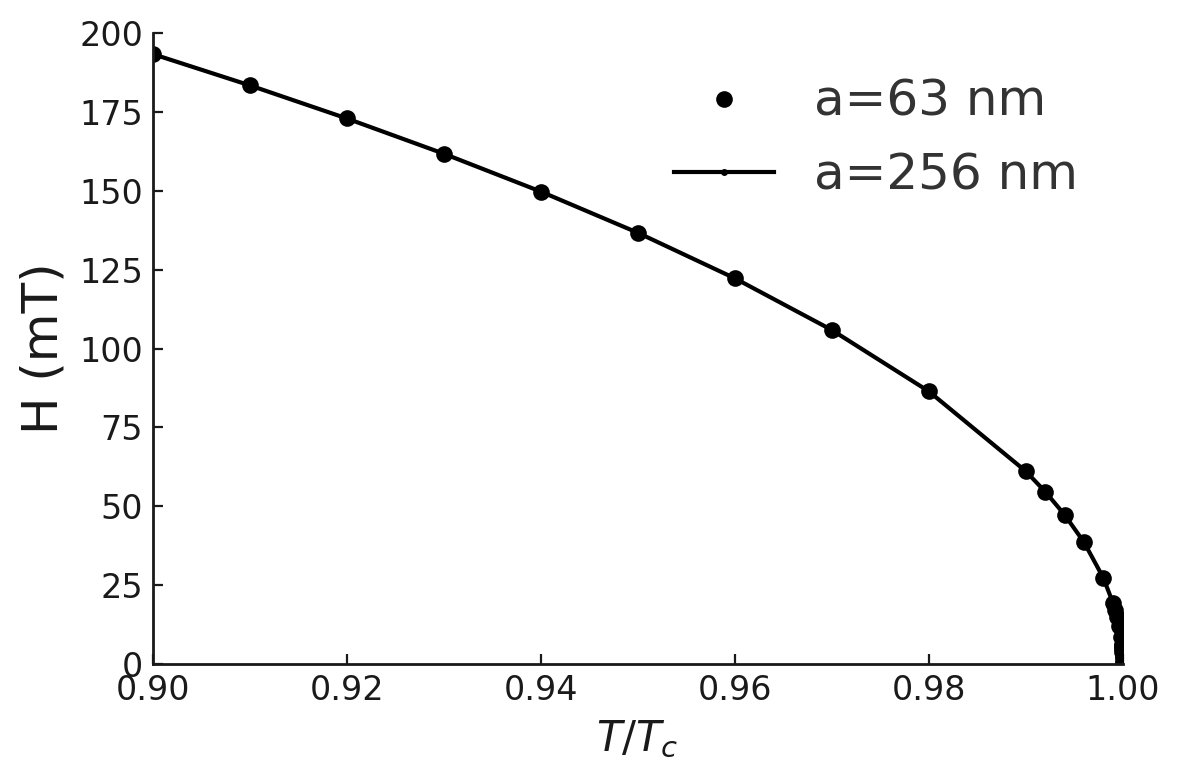}}
    \caption{Critical field for a Casimir cavity formed by a 25 nm thin Pb film and a 80 nm Au plate in a cavity with a separation of 63 nm(dots) and 256 nm(line).}
\end{figure}

%Below is my text for the critical field; above is Abhishek's
%% Figure 4 was removed from here\
%\begin{figure}[h]
%\centerline{\includegraphics[width=12.0cm]{H_final_bw.png}}
%\caption{ The critical field for a Casimir cavity formed by a 25nm thin Pb films and a 80 nm, Au plate in  cavity
%with a separation of 63 nm (points) and 256 nm (line). Figure from PhD thesis of Abhishek Som, unpublished.}
%\label{fig4}}
%\end{figure}

%
%\section{Theoretical Interpretation of the Experiment}
%\newpage

\section{Conclusions, Future Directions, and Possible Physical Implications}

In this presentation we have reviewed our development of a unique nanomechanical-transducer-based measurement technique that we have used in a careful series of experiments to directly measure the shifts in the Casimir energy by placing a superconducting Pb film in a cavity and tuning the gap, looking for effects on the superconducting transition temperature of the film. Our chip-scale system can deposit and measure a superconducting thin film while simultaneously actuating a nearby plate, forming a tunable Casimir cavity. The {\it in situ} superconducting film deposition process is achieved with two arrays of MEMS heaters that have been preloaded with a thick film of lead (Pb) and can be pulsed at low temperature to evaporate small amounts of material. The thin film is produced by using a shadow mask to define a precise pattern of the evaporated Pb (incident on the mask from two sides) that connects the four metallic measurement leads and creates a thin section of Pb directly underneath the movable gold (Au) plate. By driving a current with two sets of leads and measuring the voltage drop with the other two, we can measure the resistance of Pb. We monitor this resistance as the Au plate is driven to its mechanical resonance and back to zero amplitude. Using finite element analysis, we are able to estimate the deformed mode shape at resonance, which places the minimum separation of Pb and Au between 63 and 73 nm. The maximum separation is determined to be $\sim$ 256 nm. The difference in Casimir energy is calculated as the cavity size oscillates from minimum to maximum separation\cite{diego2020, bimonte2005}. This translates to a theoretically predicted difference in $T_c$ to be about 0.025 micro K, which is roughly 1000 times smaller than our experimental resolution of 12 micro K. In terms of future directions, we plan to add a magnetic field to our {\it in situ} apparatus to enable a direct measurement of  the effects of the magnetic field instead of our current indirect measurements. Regarding possible physical implications of the successful observation of a renormalization of the zero of the vacuum energy by the Casimir cavity, we refer the interested reader to \cite{stange2021}, where these are discussed in more detail than space here allows. But they range from applications to manipulation of nanoscale objects such as a micro-tractor for moving quantum dots, nanowires, and bacteria to highly speculative possibilities such as the possibility of stabilizing  the existence of wormholes and and an explanation of the dark energy of the universe \cite{stange2021}. The future of the Casimir effect and the Casimir energy is promising indeed.

\section*{Acknowledgements}

It is a pleasure to thank our collaborators in this work -- Alexander Stange, Lawrence Barrett, Matthias Imboden, and Vladimir Aksyuk -- and the Division of Materials Science and Energy and the High Performance Computing Center at Boston University for their support.

%\begin{thebibliography}{000} %for 3 digits

\end{document}